**Growing homophilic networks are natural navigable small worlds**

Yury A. Malkov[1], Alexander Ponomarenko[2]

1. Federal state budgetary institution of science Institute of Applied Physics of the Russian Academy of Sciences, 46 Ul'yanov Street, 603950 Nizhny Novgorod, Russia
2. National Research University Higher School of Economics, Nizhny Novgorod, Russia

Correspondence: yurymalkov@mail.ru

**Abstract.** Navigability, an ability to find a logarithmically short path between elements using only local information, is one of the most fascinating properties of real-life networks. However, the exact mechanism responsible for the formation of navigation properties remained unknown. We show that navigability can be achieved by using only two ingredients present in the majority of networks: network growth and local homophily, giving a persuasive answer how the navigation appears in real-life networks. A very simple algorithm produces hierarchical self-similar optimally wired navigable small world networks with exponential degree distribution by using only local information. Adding preferential attachment produces a scale-free network which has shorter greedy paths, but worse (power law) scaling of the information extraction locality (algorithmic complexity of a search). Introducing saturation of the preferential attachment leads to truncated scale-free degree distribution that offers a good tradeoff between these parameters and can be useful for practical applications. Several features of the model are observed in real-life networks, in particular in the brain neural networks, supporting the earlier suggestions that they are navigable.

## Introduction

Large scale networks are ubiquitous in many domains of science and technology. They influence numerous aspects of daily human life, and their importance is rising with the advances in the information technology. Even human's ability to think is governed by a large-scale brain network containing more than 100 billion neurons[1]. One of the most fascinating features found in the real-life networks is the navigability, an ability to find a logarithmically short path between two arbitrary nodes using only local information, without global knowledge of the network.

In the late 1960's Stanley Milgram and his collaborators conducted a series of experiments in which individuals from the USA were asked to get letters delivered to an unknown recipient in Boston[2]. Participants forwarded the letter to an acquaintance that was more likely to know the target. As a result about 20% of the letters arrived to the target on the average in less than six hops. In addition to revealing the existence of short paths in real-world acquaintance networks, the small-world experiments showed that these networks are navigable: a short path was discovered through using only local information. Later, the navigation feature was discovered in other types of networks[3]. The first algorithmic navigation model with a local greedy routing was proposed by J. Kleinberg[4, 5], inspiring many other studies and applications of the effect (see the recent review in [3]). However, the exact mechanism that is responsible for formation of navigation properties in real-life networks remained unknown. It was recently suggested that the navigation properties can rise due to various optimization schemes, such as optimization of network's entropy[6], optimization of network transport[7-11], game theory models[12, 13] or due to internal hyperbolicity of a hidden metric space[14]. In [15] a realistic model based on random heterogeneous networks was proposed to describe navigation processes in real networks. Authors argued that to achieve a high probability of successful routing in the large network limit the network has to have a scale-free degree distribution with $\gamma<2.5$. Later, hyperbolicity of hidden metric space was proposed as a possible reason of forming such navigable structures in real-life networks [13, 14, 16]. However it is unclear whether hyperbolicity and other aforementioned complex schemes are related to processes in real-life networks. In this work we show that the navigation property can be directly achieved by using only two ingredients that are present in the majority of real-life networks: network growth and local homophily[17], giving a simple and persuasive answer to the question of the nature of navigability in real-life systems.

One of the natural byproducts of the navigation studies is emergence of new efficient algorithms for distributed data similarity search (namely, the K-Nearest Neighbor Search, K-NN) which is a keen problem for many applications[18]. Several network structures inspired by the Kleinberg's idea were proposed[19-21]; their realization, however, was far from practical applications. In refs. [22-24] an efficient approximate KNN algorithm was introduced for general metric data utilizing incremental insertion and connection of newcoming elements to their closest neighbors in order to construct a navigable small world graph. By simulations the authors show that the algorithm can produce networks with short greedy paths and achieves a polylogarithmic complexity for both search and insertion firmly outperforming rival algorithms for a wide selection of datasets[24-26]. However, the scope of the works[22-24] was limited to the approximate nearest neighbors problem.

Based on these ideas we propose Growing Homophilic (GH) networks as the origin of small world navigation in real-life systems. We analyze the network properties using simulations and theoretical consideration, confirming navigation properties and demonstrating that the scale-free navigation models[15, 27] considered earlier are not truly local in terms of information extraction locality (algorithmic complexity of a search), while the proposed model is. We also show that the GH network features can be found in real-life networks, with an emphasis on functional brain networks.

Functional brain networks are studied in vivo using MRI techniques[28] and are usually modeled by generalizations of random models[29-31] requiring global network knowledge. It was suggested that the brain networks are navigable through utilizing the rich club (a densely interconnected high degree subgraph[32]) and that the navigation plays a major role in brain's function[33]. In the recent work[13] it was demonstrated that the functional brain networks have a navigation skeleton that allows greedy searching with low errors. Both growth and homophily[30, 34] are usually considered to be important factors influencing the brain network structure. Local connection to nearby neurons together with network growth are considered as a plausible mechanism for formation of long range connections in small nervous networks[35, 36], similarly to the proposed model. Our study shows that the GH networks have high level features that are found in the functional brain networks, indicating that the GH mechanism is not suppressed and plays a significant role in brain network formation, thus supporting the earlier suggestions that the brain networks are naturally navigable.

GH networks can also be viewed as a substantial generalization of a complex growing spatial 1D OHO model introduced in [37], which was used to deterministically produce networks with high clustering and short average path. Recently, another generalization of this model for the multidimensional case was proposed as a possible mechanism for formation of neural networks[38]. However, formation of navigation properties was not a subject of studies of the OHO model. GH network in the case of 1D circle data can be also considered as a degenerated version of growth models studied in [16] with an exclusion of popularity term (which also makes it similar to the OHO model). It was demonstrated that the hyperbolic model from [16], which is a growing model in a hyperbolic space, adequately describes evolution of many scale-free real networks. However, the properties for the case without hyperbolicity (popularity) which leads to exponential degree distribution were studied poorly. As follows from the navigation models in refs. [14, 15], without the scale-free degree distribution such networks were not expected to be navigable.

## Results

**Construction and navigation properties**. To construct a GH network we use a set of elements *S* from a metric space $\sigma$ and a single construction parameter *M*. We start building network by inserting a random element from *S*. Then we iteratively insert randomly selected remaining elements *e* by connecting to *M* the previously inserted elements that have minimal distance $\Delta$ to *e*, until all elements from *S* are inserted. Unlike the models from refs. [4, 5, 7, 8, 15, 16, 37-40] and the Watts-Strogatz model [41] (which all require global network knowledge at construction), the GH algorithm insertions can be done approximately using only local information by selecting the approximate nearest neighbors through a help of network navigation feature (see Methods section for details). This has a clear interpretation: new nodes in many real networks do not have global knowledge, so they have to navigate the network in order to find their place and adapt. The tests showed that under appropriate parameters there is no measurable difference in network metrics whether the construction had exact or inexact neighbors selection, while the network assembly process was drastically faster in the approximated neighbors case.

Because the elements from *S* are not placed on a regular lattice, the greedy search algorithm can be trapped in a local minimum before reaching the target. The generalization of the regular lattice for this case is the Delaunay graph, which is dual to the Voronoi partition. If we have a Delaunay graph subset in the network, the greedy search always ends at an element from *S* which is closest to any target element $t \in \sigma$ [42] (note that this condition is more general than the one studied in [13]). It is easy to construct a Delaunay graph for low dimensional Euclidian spaces, especially in 1D case where it is a simple liked list, however it was shown that constructing the graph using only distances between the set elements is impossible for general metric spaces[42]. Still, it was demonstrated that connecting to *M* nearest neighbors acts as a good enough approximation of the Delaunay graph, so that by increasing *M* or using a slightly modified versions of the greedy algorithm these effects can be made negligible[23, 24]. The average greedy algorithm hop count of a GH network for different input data is presented in fig. 1(a). The graph shows a clear logarithmic scaling for all data used, including a non-trivial case of edit distance for English words. At the parameters used the probability of a success navigation was higher than 0.92 for all the data and higher than 0.999 for vector data with d<5.

The definition of navigability in [15] requires also that the greedy search success probability does not tend to zero in the limit of infinite network size. This restriction has led to a conclusion that the navigability should be expected only for the power law degree distribution networks with $\gamma$<2.5. However for the 1D data case (which was the only one considered by the authors of [15]) the success probability can be fixed to unity by a very simple realistic heuristics, thus questioning such a conjecture.

Networks produced by a GH model are also navigable by the aforementioned definition at least at small dimensionality (d≤6), as tests indicate that the recall converges to a constant value (see Supporting information for details), thus demonstrating existence of a new class of navigable models.

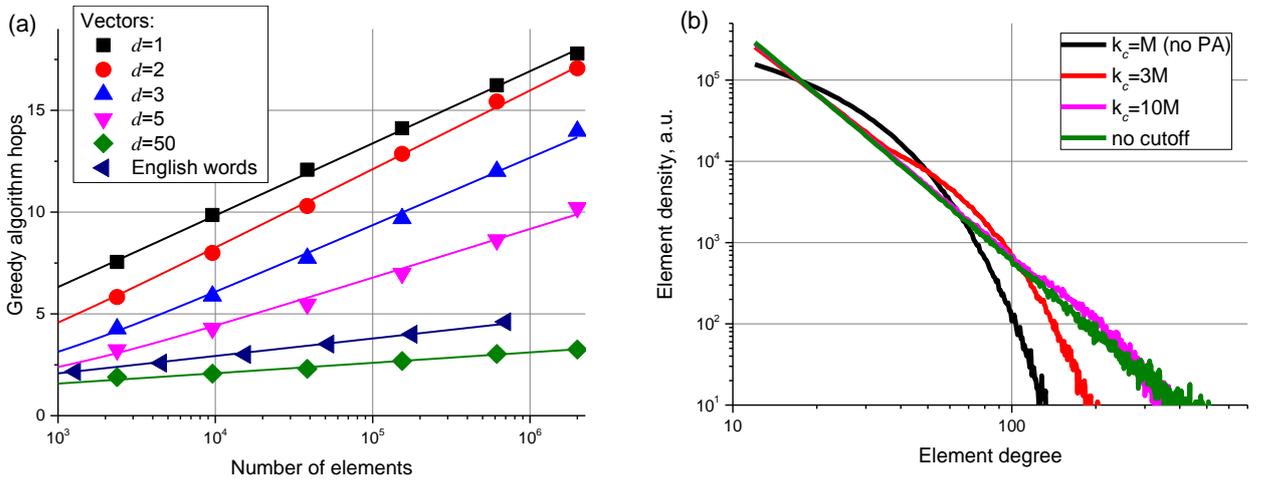

Fig. 1. (a) Average hop count during a greedy search for different dimensionality Euclidian data and English words database with edit distance showing logarithmic scaling. (b) Degree distribution for the GH algorithm networks with PA for different degree cutoffs ($k_c$).

**Degree distribution and information extraction locality.** A GH network has an exponential degree distribution (see in fig. 1(b)). The scale in the exponent is determined by the *M* parameter, similar analysis can be done as in [37, 38]. The exponential degree distribution is present in the real-life networks such as power grids, air traffic networks, and collaboration networks of company directors[43]. Studies of the functional brain network degree distribution yielded ambiguous results: some investigations have shown exponential degree distribution, while others exhibited scale-free or truncated scale-free distribution[44].

Most of the studied real-life networks, however, have a power law degree distribution. The GH algorithm can be slightly modified by adding a preferential attachment (PA)[45] to produce a scale-free (power law) degree distribution (which makes it similar to the growing models in a hyperbolic plane[16]). To achieve that, the distances to the elements are normalized by $k^{1/d}$ during the network construction for uniform data in Euclidean space (see the Methods section for the details), leading to a power law degree distribution with $\gamma$ close to 3. With the addition of a cutoff $k_c$, the degree distribution transforms into a power law with an exponential cutoff (see fig. 1(b)). As expected for scale-free networks[15], adding the preferential attachment does not suppress the network navigability (see fig. S1(b) in Supporting information for the comparison).

Exponential degree distribution is usually attributed to limited capacity of a node or to absence of PA mechanisms[41]. However, there is another critical distinction between scale-free and exponential degree distributions in terms of locality of information extraction which arises in virtual computer networks having practically no limit on node capacity. We define the locality as the number of distance computations during a greedy search, which also corresponds to algorithmic complexity of a search algorithm. Our simulations show that, for the scale-free networks (both in GH networks with PA and scale-free networks studied in [15]), the number of distance computations has a *power law* scaling with the number of network elements, in contrast to GH networks without PA and Kleinberg's networks which have a *polylogarithmic* scaling[24](see fig. S1 and fig. S2 in Supporting information). This happens because the greedy algorithm prefers nodes with the highest degrees (which have monopoly on long range links)[15] while the maximum degree in scale-free network has power law scaling $N^{1/(\gamma-1)}$ with the number of elements[46] leading to $N^{1/(\gamma-1)}$ search complexity. The authors in [14] argued that the best choice for optimal navigation is when $\gamma$ is close to 2; this, however, leads to almost linear scaling of greedy search distance computations number. Such scaling makes using scale-free networks impractical for greedy routing in large-scale networks where high locality of information extraction matters, which is the case of K-NN algorithms and is likely to be the case for the brain networks.

The importance of employing the locality of information extraction as a measure for network navigation studies can be underpinned by considering a star graph (a graph where every node is connected to a single central hub) with a modified greedy search algorithm that uses effective distance $\Delta \cdot e^{ak}$ (where $k$ is degree of a candidate node, $a$ is a parameter, $\Delta$ is initial metric distance between the candidate and the target). The parameter $a$ can be always set large enough, so that every greedy path will go through the hub reaching the target in two steps regardless of the network size. This network is ideal in all navigation measures defined in [14, 15], having short path, ideal success ratio and low average degree. However, to find these paths the greedy search algorithm has to utilize the hub's global view of the network and to compare distances to every network node thus having a bad linear algorithmic complexity scaling.

At the same time, the scale-free networks offer less greedy algorithm hops compared to the base GH algorithm which is beneficial. A power law degree distribution with an exponential cutoff seems to be a good tradeoff between low number of hops and low complexity of a search. Slightly increasing the cutoff $k_c$ above $M$ in GH algorithm with PA sharply decreases the number of greedy algorithm hops (see fig. S1(b)), while having almost no impact on the number of distance computations (fig. S1(a)). This finding can be used for constructing artificial networks optimized for best navigability both in terms of complexity and number of hops.

**Link length distribution and optimal wiring.** For uniformly distributed bounded $d$-dimensional Euclidian data, the average distance between the nearest neighbors scales as $r(N) \sim N^{-1/d}$ (1) with the total number of elements $N$ in the network. It means that every characteristic scale of link length in a final network can be put in correspondence to some specific time of construction. That allows deducing the link length distribution by differencing the equation (1). By doing this we get a power law link length density $dN \sim r^{-\alpha}dr$ with $\alpha = d+1$ exponent (confirmed by the simulations). It was recently shown[7, 8] that $\alpha = d+1$ is the optimal value for the shortest path length and greedy navigation path in case of constraint on total length of all connections in the network. Thus, GH networks are naturally close to optimal in terms of the wiring cost.

Power law link length distributions with $\alpha = d+1$ are encountered in real-life networks like airport connections networks[47] and functional brain networks[48, 49]. It was speculated in refs. [7, 8] that such behavior arises due to global optimization schemes, while GH networks provide a much more simple and natural explanation for the exponent value.

**Self-similarity and hierarchical modular structure**. Construction of a GH network is an iterative process: at each step we have as an input a navigable small world network and we insert new elements and links preserving its properties. A part of a uniform data GH network covered by a ball is also a navigable small world with few outer connections. Thus, the GH networks have self-similar structure. Analysis of self-similarity identical to [27] is presented in the Supporting Information in fig. S3, demonstrating a self-similar structure of network's clustering coefficient.

The hierarchical self-similarity property is found in many real-life networks[27, 50, 51]. Studies have shown that the functional brain networks form a hierarchically modular community structure[48, 52] consisting of highly interconnected specialized modules, only loosely connected to each other. This may seem to contradict the small world feature which is usually modeled by random networks[48]. GH networks can easily model both small world navigation and modular structure simultaneously by introducing clusters. In this case, coordinates of the cluster centers may correspond to different neuron specialties in a generalized underlying metric space. A 2D GH network for clustered data is presented in fig. 3 demonstrating highly modular and at the same time navigable network structure containing only 30 intermodular links (0.03% of the total number) between the first elements in the network (which form a rich club). By using a simple modification of the greedy algorithm with preference of high degree nodes (see Methods section) short paths between different module elements can be efficiently found using only local information.

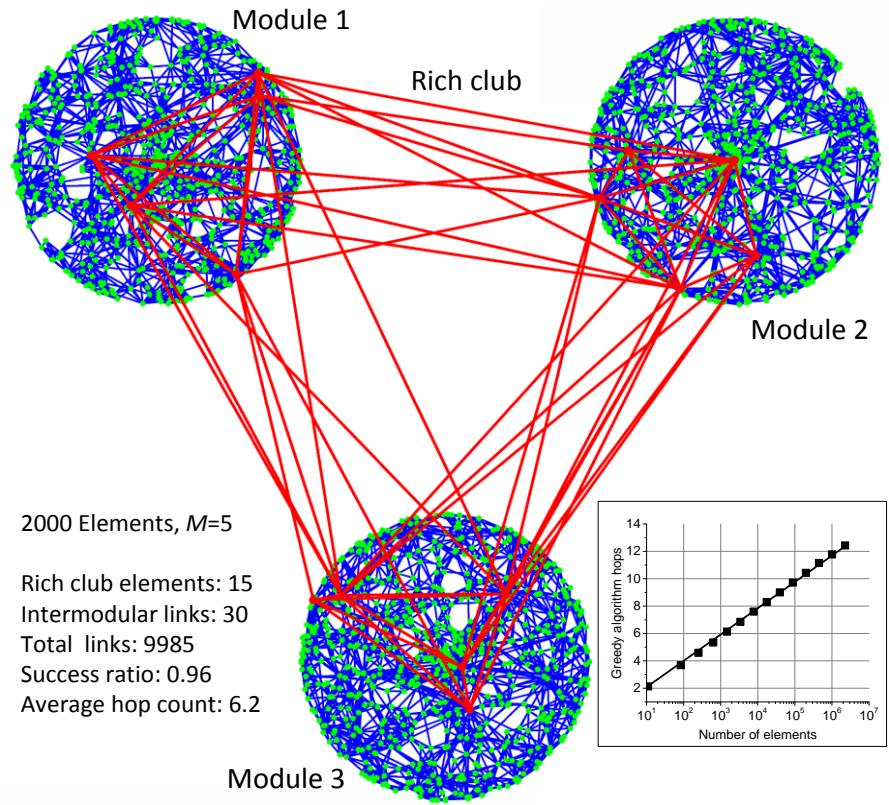

2000 Elements, $M$=5

Rich club elements: 15
Intermodular links: 30
Total links: 9985
Success ratio: 0.96
Average hop count: 6.2

Fig. 2. 2D network constructed by the GH algorithm with $M$=5 for clustered $d$=2 Euclidian data. The inset shows scaling of the modified greedy algorithm average hop count.

**Rich club and greedy hops upper bounds.** Simulations show that probability of a connection between the network elements grows exponentially with the element degree, thus demonstrating the presence of a rich club (see fig. S5 in Supporting Information). Due to incremental construction and self-similarity of the GH networks every preceding instance of a GH network acts as a rich club to any subsequent instance. To achieve a well-defined rich club, the self-similarity symmetry has to be broken by introducing non-fractality in data, such as a fixed number of clusters in fig. 2.

Universally, the rich club is composed of the first elements inserted by the GH algorithm, which is also the case for the brain networks. Studies have shown that the rich club in human brain is formed before the 30[th] week of gestation with almost no changes of its inner connections until birth[53]. Moreover, the investigations of C. elegans worms neural network have shown that the rich club neurons are among the first neurons to be born[54, 55]. Thus, together with [35], the GH model offers a plausible explanation of how the rich clubs are formed in brain networks.

Due to presence of rich clubs in GH networks a general navigation analysis similar to the scale-free networks from [15] can be done. At the beginning of a greedy search the algorithm "zooms-out" preferring high degree nodes with a higher characteristic link radius until it reaches a node for which the characteristic radius of the connections is comparable with the distance to the target node. Next, a reverse "zoom-in" procedure takes place until the target node is reached, see [15] for details.

We offer a slightly different perspective. It can be shown that in GH networks the rich club is also *navigable*, meaning that a greedy search between two rich hub nodes is very likely to select only the rich club nodes at each step. This is illustrated by the simulations results in fig. 3(a) showing that the average hop count for the first $10^4$ elements selected as start and targets nodes does not depend on the dataset size, i.e. the greedy search algorithm ignores newly added links. Figure 3(b) shows a schematic Voronoi partition of rich club element connections for a greedy algorithm step with another rich club element as a target. In the case of a good Delaunay graph approximation (high enough *M*) the Voronoi partition alters only locally as is shown in fig. 3(b), thus having no impact on the greedy search between the rich club elements.

Self-similarity and navigability of the rich club in GH networks play a crucial role in the navigation process. Under the conditions of a perfect Delaunay graph and rich club navigability one can show that there is an upper greedy algorithm hop bound $2\log_2(N)$ (see Supporting Information for details), thus proving a logarithmic greedy search hops scaling.

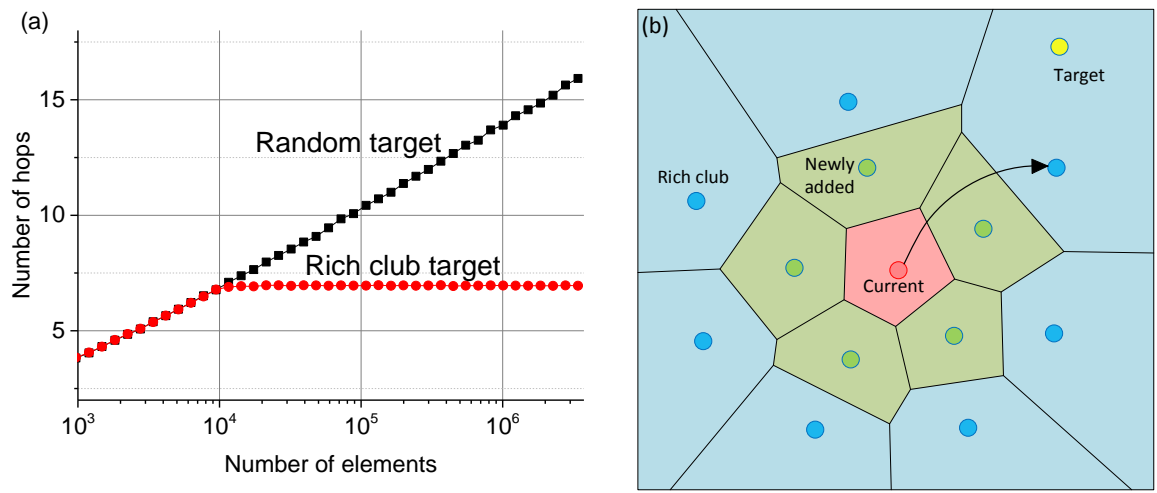

Fig. 3 (a) Average number of greedy algorithm hops scaling for the first $10^4$ elements given as start and target nodes (red) and all elements used for search (black). The first $10^4$ elements form a rich club that ignores more newly added elements. The results are presented for Euclidian data with *d*=2, *M*=20.
(b) Cartoon of Voronoi partition for connections of a single greedy search step. Newly added elements (green) cause only local changes in Voronoi partitioning, so if the target element lies outside the current element connections, it falls into Voronoi partition of rich club's elements (blue), thus ignoring local connections at greedy search.

## Discussion

Using simulations and theoretical studies we have demonstrated that two ingredients that are present in the majority of networks, namely network growth and local homophily, are sufficient to produce a navigable small world network, giving a simple and persuasive answer how the navigation feature appears in real-life networks. In contrast to the generally used models, a simple local GH model without central regulation by using only local information produces hierarchical self-similar optimally wired navigable networks, which offers a simple explanation why these features are found in real-life networks without a need of employing hyperbolicity or other complex schemes. Self-similarity and rich clubs navigation of the GH networks lead to emergence of logarithmic scaling of the greedy algorithm hops in GH networks.

By adding PA with saturation the degree distribution can be tuned from exponential to scale-free with or without exponential cutoff. We have shown that in case of pure scale-free degree distribution (as well as for the scale-free networks studied in [15] and hyperbolic networks), the true logarithmic local routing cannot be achieved due to a power law scaling of information extraction locality (algorithmic complexity of a search). Truncated scale-free degree distribution offers a reasonable tradeoff between the path length and the algorithmic complexity and can be used for practical applications.

Thus, every network that has both growth and homophily is a potential navigable small-world network. This is an important finding for real-life networks, as real life is full of examples of growth and homophily shaping the network. Scientific papers form a growing navigable citation network (navigability of which is actively utilized by researchers) just by citing existing related works. Big city passenger airports were among the first to open and later became big network hubs which play an important role in spatial navigation in airport networks[15]. Neural brain networks are formed utilizing both growth and homophily, producing hierarchical structures with rich clubs consisting of early born neurons. The proposed GH model offers a conclusive explanation of navigability in these networks. Still, the evolution of some networks including social structures incorporate other factors, such as node moving and departure. For these networks, a sudden departure of a major node (say, a key manager in a company or rich club neurons[56]) can seriously hurt the performance. However, some of these networks can be resilient to the above-mentioned processes, thus preserving navigation. For example, there are quite a few top-level deputies in big companies, and when a manager quits, he/she has to pass his/her contacts to a newcomer.

There is evidence that in real-life networks such as an airport and brain networks, the GH model is not suppressed by other mechanisms. In addition to the aforementioned growth and homophily, several other high level features of the GH model are observed in brain networks, such as low diameter, high clustering, presence of navigation skeleton, hierarchical self-similar modular structure, power law link length distribution with exact $d+1$ exponent, and emergence of a rich club from the first elements in the network. This indicates that the model plays a significant role in the formation of brain networks and that they are likely to be navigable, supporting the earlier suggestions[13, 33].

The proposed GH model can be used as a guide for building artificial optimally wired navigable structures using only local information.

## Methods

**Construction.** The GH network was constructed through iteratively inserting the elements into the network in random order by adding bidirectional links to the *M* closest elements. To find the connections we applied approximate K-NN graph algorithms[24] (C++ implementations of the K-NN algorithms are available in the Non-Metric Space Library[57]) which had utilized the navigation in the constructed graph. To obtain approximate *M* nearest neighbors, a dynamic list of *M* closest of the found elements (initially filled with a random enter point node) was kept during the search. The list was updated at each step by evaluating the neighborhood of the closest previously non-evaluated element in the list until the neighborhood of every element from the list was evaluated. For *M*=1, this method is equivalent to a basic greedy search. The best *M* results from several trials were used as approximate closest elements. The number of trials was adjusted so that the recall (the ratio between the found and the true *M* nearest neighbors) was higher than 0.95, producing results almost indistinguishable from what you get from the exact search. The construction procedure has a polylogarithmic complexity[24].

For the simulations with PA (fig. 1(a)), effective distances to the elements ($\Delta_{eff}(e1,e2)=\Delta_{Euclid}(e1,e2)/(k_2)^{1/d}$ if $k_2<k_c$, $\Delta_{new}(e1,e2)=\Delta_{Euclid}(e1,e2)/(k_c)^{1/d}$ otherwise, *k* is the degree of a node) were used to select the neighbors. Changing the seed of the algorithm random data generators, connecting to the *M* exact neighbors and/or construction in many parallel threads had a very slight effect on the evaluated network metrics.

**Datasets.** Random Euclidian data with coordinates distributed uniformly in [0,1] range with $L_2$ distance was used to model the vectors. For testing the Damerau–Levenshtein distance, about 700k English words from the Scowl Debain database were used as the dataset. In order to unambiguously select the next node during a greedy search, a small random value was added to the Damerau–Levenshtein distance. For fig. 1(a) parameter *M* was set to be 9, 12, 20, 25, 150 and 40 for Euclidian vectors with *d*=1, 2, 3, 5, 50 and English word dataset, respectively. The success ratio for the vectors was higher than 0.999 for *d*≤5, higher than 0.92 for *d*=50 and English words data. On the average, only 760 distance computations were needed to find the path between two arbitrary words in 700k database and only about 1280 distance computations were required to find the path for 20 million *d*=5 Euclidian vectors.

**Network metrics.** To evaluate the average number of hops, we used up to $10^4$ randomly selected nodes as start and target elements. The greedy algorithm selects at each step a neighbor that is closest to the target as an input for the next step, until it reaches the element which is closer to the target than its neighbors. The success ratio is the ratio of the number of successful searches to the total number of searches. The search is considered failed if the result is not the target element.

The information extraction locality metric was evaluated by counting the average number of distance calculations during a single greedy search.

To get a high recall (>0.95) for the tests with clustered 2D data and for the e we used a modification of the greedy search algorithm[39] that minimized the normalized distance ($\Delta_{norm}(target,e2)=\Delta_{Euclid}(target,e2)/(k_2)^{1/2}$) .

### Acknowledgements
We are grateful to I. Malkova, L. Boytsov, D. Yashunin, N. Krivatkina and V. Nekorkin for the discussions. The reported study was funded by RFBR, according to the research project No. 16-31-60104 mol_a_dk.

### References
1. Arbib MA. The handbook of brain theory and neural networks: MIT press; 2003.
2. Travers J, Milgram S. An experimental study of the small world problem. Sociometry. 1969:425-43.
3. Huang W, Chen S, Wang W. Navigation in spatial networks: A survey. Physica A: Statistical Mechanics and its Applications. 2014;393:132-54.
4. Kleinberg JM. Navigation in a small world. Nature. 2000;406(6798):845-.
5. Kleinberg J, editor The small-world phenomenon: An algorithmic perspective. Proceedings of the thirty-second annual ACM symposium on Theory of computing; 2000 2000: ACM.
6. Hu Y, Wang Y, Li D, Havlin S, Di Z. Possible origin of efficient navigation in small worlds. Physical Review Letters. 2011;106(10):108701.
7. Li G, Reis S, Moreira A, Havlin S, Stanley H, Andrade Jr J. Optimal transport exponent in spatially embedded networks. Physical Review E. 2013;87(4):042810.
8. Li G, Reis S, Moreira A, Havlin S, Stanley H, Andrade Jr J. Towards design principles for optimal transport networks. Physical Review Letters. 2010;104(1):018701.
9. Chaintreau A, Fraigniaud P, Lebhar E. Networks become navigable as nodes move and forget. Automata, Languages and Programming: Springer; 2008. p. 133-44.
10. Clarke I, Sandberg O, Wiley B, Hong TW, editors. Freenet: A distributed anonymous information storage and retrieval system. Designing Privacy Enhancing Technologies; 2001: Springer.
11. Sandberg O, Clarke I. The evolution of navigable small-world networks. arXiv preprint cs/0607025. 2006.

12. Yang Z, Chen W, editors. A Game Theoretic Model for the Formation of Navigable Small-World Networks. Proceedings of the 24th International Conference on World Wide Web; 2015: International World Wide Web Conferences Steering Committee.
13. Gulyás A, Bíró JJ, Kőrösi A, Rétvári G, Krioukov D. Navigable networks as Nash equilibria of navigation games. Nature Communications. 2015;6:7651.
14. Krioukov D, Papadopoulos F, Kitsak M, Vahdat A, Boguná M. Hyperbolic geometry of complex networks. Physical Review E. 2010;82(3):036106.
15. Boguna M, Krioukov D, Claffy KC. Navigability of complex networks. Nature Physics. 2009;5(1):74-80.
16. Papadopoulos F, Kitsak M, Serrano MÁ, Boguñá M, Krioukov D. Popularity versus similarity in growing networks. Nature. 2012;489(7417):537-40.
17. McPherson M, Smith-Lovin L, Cook JM. Birds of a feather: Homophily in social networks. Annual Review of Sociology. 2001:415-44.
18. Chávez E, Navarro G, Baeza-Yates R, Marroquín JL. Searching in metric spaces. ACM computing surveys (CSUR). 2001;33(3):273-321.
19. Beaumont O, Kermarrec A-M, Marchal L, Rivière É, editors. VoroNet: A scalable object network based on Voronoi tessellations. Parallel and Distributed Processing Symposium, 2007 IPDPS 2007 IEEE International; 2007: IEEE.
20. Beaumont O, Kermarrec A-M, Rivière É. Peer to peer multidimensional overlays: Approximating complex structures. Principles of Distributed Systems: Springer; 2007. p. 315-28.
21. Lifshits Y, Zhang S, editors. Combinatorial algorithms for nearest neighbors, near-duplicates and small-world design. Proceedings of the Twentieth Annual ACM-SIAM Symposium on Discrete Algorithms; 2009: Society for Industrial and Applied Mathematics.
22. Ponomarenko A, Mal'kov Y, Logvinov A, Krylov V, editors. Approximate Nearest Neighbor Search Small World Approach. International Conference on Information and Communication Technologies & Applications; 2011; Orlando, Florida, USA.
23. Malkov Y, Ponomarenko A, Logvinov A, Krylov V. Scalable distributed algorithm for approximate nearest neighbor search problem in high dimensional general metric spaces. Similarity Search and Applications: Springer Berlin Heidelberg; 2012. p. 132-47.
24. Malkov Y, Ponomarenko A, Logvinov A, Krylov V. Approximate nearest neighbor algorithm based on navigable small world graphs. Information Systems. 2014;45:61-8.
25. Ponomarenko A, Avrelin N, Naidan B, Boytsov L. Comparative Analysis of Data Structures for Approximate Nearest Neighbor Search. In Proceedings of The Third International Conference on Data Analytics. 2014.
26. Naidan B, Boytsov L, Nyberg E. Permutation search methods are efficient, yet faster search is possible. VLDB Procedings. 2015;8(12):1618-29.
27. Serrano MA, Krioukov D, Boguná M. Self-similarity of complex networks and hidden metric spaces. Physical Review Letters. 2008;100(7):078701.
28. Bullmore E, Sporns O. Complex brain networks: graph theoretical analysis of structural and functional systems. Nature Reviews Neuroscience. 2009;10(3):186-98.
29. Vértes PE, Alexander-Bloch A, Bullmore ET. Generative models of rich clubs in Hebbian neuronal networks and large-scale human brain networks. Philosophical Transactions of the Royal Society of London B: Biological Sciences. 2014;369(1653):20130531.
30. Betzel RF, Avena-Koenigsberger A, Goñi J, He Y, de Reus MA, Griffa A, et al. Generative models of the human connectome. NeuroImage. 2015.
31. Vértes PE, Alexander-Bloch AF, Gogtay N, Giedd JN, Rapoport JL, Bullmore ET. Simple models of human brain functional networks. Proceedings of the National Academy of Sciences. 2012;109(15):5868-73.
32. Colizza V, Flammini A, Serrano MA, Vespignani A. Detecting rich-club ordering in complex networks. Nature Physics. 2006;2(2):110-5.
33. van den Heuvel MP, Kahn RS, Goñi J, Sporns O. High-cost, high-capacity backbone for global brain communication. Proceedings of the National Academy of Sciences. 2012;109(28):11372-7.
34. Bullmore E, Sporns O. The economy of brain network organization. Nature Reviews Neuroscience. 2012;13(5):336-49.

35. Lim S, Kaiser M. Developmental time windows for axon growth influence neuronal network topology. Biological cybernetics. 2015;109(2):275-86.
36. Nicosia V, Vértes PE, Schafer WR, Latora V, Bullmore ET. Phase transition in the economically modeled growth of a cellular nervous system. Proceedings of the National Academy of Sciences. 2013;110(19):7880-5.
37. Ozik J, Hunt BR, Ott E. Growing networks with geographical attachment preference: Emergence of small worlds. Physical Review E. 2004;69(2):026108.
38. Zitin A, Gorowara A, Squires S, Herrera M, Antonsen TM, Girvan M, et al. Spatially embedded growing small-world networks. Scientific Reports. 2014;4.
39. Thadakamalla H, Albert R, Kumara S. Search in spatial scale-free networks. New Journal of Physics. 2007;9(6):190.
40. Zuev K, Boguñá M, Bianconi G, Krioukov D. Emergence of Soft Communities from Geometric Preferential Attachment. Scientific Reports. 2015;5.
41. Watts DJ, Strogatz SH. Collective dynamics of 'small-world'networks. Nature. 1998;393(6684):440-2.
42. Navarro G. Searching in metric spaces by spatial approximation. The VLDB Journal. 2002;11(1):28-46.
43. Newman M, Barabasi A-L, Watts DJ. The structure and dynamics of networks: Princeton University Press; 2006.
44. Qi S, Meesters S, Nicolay K, Romeny BMtH, Ossenblok P. The influence of construction methodology on structural brain network measures: A review. Journal of Neuroscience Methods. (0). doi: http://dx.doi.org/10.1016/j.jneumeth.2015.06.016.
45. Barabási A-L, Albert R. Emergence of scaling in random networks. Science. 1999;286(5439):509-12.
46. Boguná M, Pastor-Satorras R, Vespignani A. Cut-offs and finite size effects in scale-free networks. The European Physical Journal B-Condensed Matter and Complex Systems. 2004;38(2):205-9.
47. Bianconi G, Pin P, Marsili M. Assessing the relevance of node features for network structure. Proceedings of the National Academy of Sciences. 2009;106(28):11433-8.
48. Gallos LK, Makse HA, Sigman M. A small world of weak ties provides optimal global integration of self-similar modules in functional brain networks. Proceedings of the National Academy of Sciences. 2012;109(8):2825-30.
49. Eguiluz VM, Chialvo DR, Cecchi GA, Baliki M, Apkarian AV. Scale-free brain functional networks. Physical Review Letters. 2005;94(1):018102.
50. Song C, Havlin S, Makse HA. Self-similarity of complex networks. Nature. 2005;433(7024):392-5.
51. Colomer-de-Simón P, Serrano MÁ, Beiró MG, Alvarez-Hamelin JI, Boguñá M. Deciphering the global organization of clustering in real complex networks. Scientific Reports. 2013;3.
52. Meunier D, Lambiotte R, Bullmore ET. Modular and hierarchically modular organization of brain networks. Frontiers in neuroscience. 2010;4.
53. Ball G, Aljabar P, Zebari S, Tusor N, Arichi T, Merchant N, et al. Rich-club organization of the newborn human brain. Proceedings of the National Academy of Sciences. 2014;111(20):7456-61.
54. Towlson EK, Vértes PE, Ahnert SE, Schafer WR, Bullmore ET. The rich club of the C. elegans neuronal connectome. The Journal of Neuroscience. 2013;33(15):6380-7.
55. Varier S, Kaiser M. Neural development features: Spatio-temporal development of the Caenorhabditis elegans neuronal network. PLOS Computational Biology. 2011.
56. Rubinov M. Schizophrenia and abnormal brain network hubs. Dialogues in Clinical Neuroscience. 2013;15(3):339.
57. Boytsov L, Naidan B. Engineering Efficient and Effective Non-metric Space Library. Similarity Search and Applications: Springer; 2013. p. 280-93. https://github.com/searchivarius/nmslib

# Supporting Information for

## "Growing homophilic networks are natural navigable small worlds"
### by Yu. A. Malkov and A. Ponomarenko

**1. Comparison of the basic GH model to the GH model with PA and to the scale-free networks from ref. [1]** (Boguna, M., Krioukov, D., & Claffy, K. C. (2009). *Nature Physics*, *5*(1), 74-80).

To evaluate the locality of information extraction we determined an average number of distance computations during a single greedy search, a measure which is usually adopted to practically determine the efficiency of similarity search algorithms. Our simulations show that for the scale-free networks produced by GH with PA the number of distance computations has a power law scaling with the number of network elements, in contrast to the GH networks without PA which have a polylogarithmic scaling (see fig. S1(a)).

At the same time, the scale-free networks offer less greedy algorithm hops compared to the base GH algorithm which is beneficial. A power law degree distribution with an exponential cutoff seems to be a good tradeoff between low number of hops and low complexity of a search. Slightly increasing the cutoff $k_c$ above $M$ in GH algorithm with PA sharply decreases the number of greedy algorithm hops (see fig. S1(b)), while having almost no impact on the number of distance computations (fig. S1(a)). This finding can be used for constructing artificial networks optimized for best navigability both in terms of complexity and number of hops.

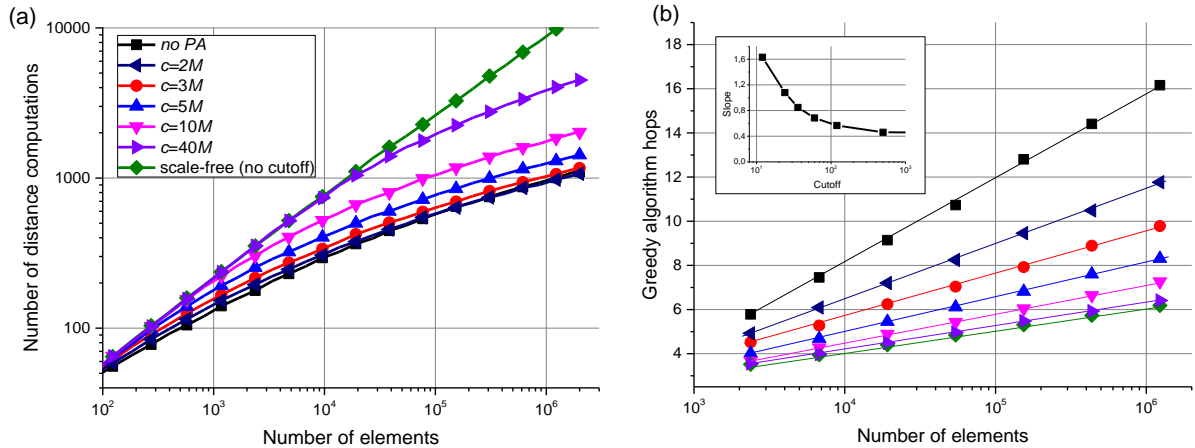

Fig. S1. Comparison of the GH model to the GH model with PA.
(a) Number of distance computations per a greedy search for GH network with PA for different degree cutoffs ($k_c$). (b) Average hop count during a greedy search for GH network with PA for different degree cutoffs. The inset shows the decay of greedy hop slope with an increase of the degree cutoff. Both plots are presented for Euclid data with $d$=2, $M$=12.

The scale-free networks from [1] were modeled with the parameters γ=5, α=2.5 as advised in the paper, for 1D uniformly distributed data in Euclidian distance. The characteristic scale parameter was adjusted to produce a success probability close to 0.92, resulting in an average node degree close to 12. The parameter *M* of the GH algorithm was set to 5, thus the network had an average degree of 10. The comparison was done only for the 1D case, because for the higher dimensions we were not able to simultaneously get high success probability and the low average degree for the scale-free networks from [1]. The maximum data size in the comparison was limited by the algorithm from [1] which had a high $N^2 \ln(N)$ complexity to establish the scaling (compared to ~$N \cdot \ln^2(N)$ complexity for the GH algorithm case). Both for the GH networks and for the scale-free networks the plots were smoothed over 8-12 trials. An averaged success probability of a greedy search does not decrease with the number of elements in the network for both models (see fig. S2(b)).

Smoothed greedy search success probability scaling is plotted in fig. S2(d) for a case of GH networks with higher dimensional data (d=4, 6). The plot shows almost no changes in the success probability as the networks grows by several orders of magnitude thus demonstrating the GH networks are also navigable by the definition from ref. [1]. This shows the existence of a new class of navigable models with exponential degree distribution. Navigability in this case was not expected from the results of ref. [1], where a scale-free degree distribution was required for the navigability.

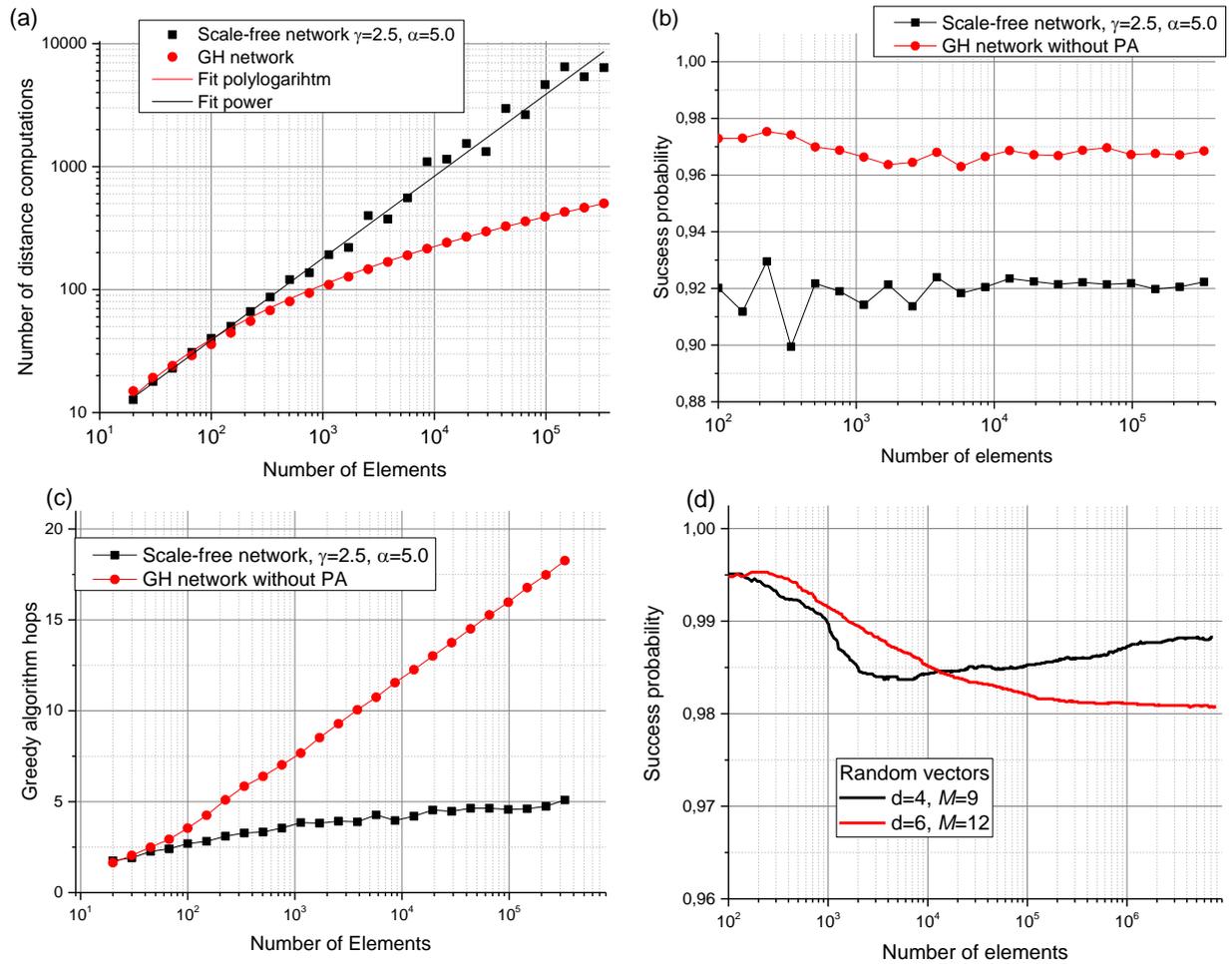

Fig S2. Comparison of the GH algorithm to the navigable scale-free networks from [1] on 1D data (a-c). (a) Average number of distance computations during a greedy search. (b) Average success probability of a greedy search. (c) Average greedy algorithm hops.
(d) The scaling of greedy search success probability in GH networks on higher dimensional random data.

While the considered scale-free networks offer significantly less hops (fig. S2(c)), the number of distance computations for the scale-free network has a scaling close to $N^{1/1.5}$ (fig. S2(a)) which corresponds to a $N^{1/(\gamma-1)}$ scaling of the maximum degree in the network. The GH networks instead have a polylogarithmic scaling. So despite the small number of hops the greedy search can hardly be called local for the scale-free networks, especially in the case of the $\gamma$ close to 2. The observed polynomial scaling should also be valid for hyperbolic geometric graphs[2] which share basic properties with networks, studied in [1].

**2. Self-similarity in clustering coefficient distribution**

The plot in fig. S3 demonstrates a self-similar structure of the network's average clustering coefficient distribution obtained using the same procedure as in [3]. All nodes with degree less than $k_{thr}$ are removed from the network and degrees of the nodes are normalized by the mean degree. Finally, a distribution of normalized average clustering coefficient is calculated for different $k_{thr}$.

Alternating the $k_{thr}$ in a wide range does not change the normalized average clustering coefficient distribution, which means that the network has a self-similar structure.

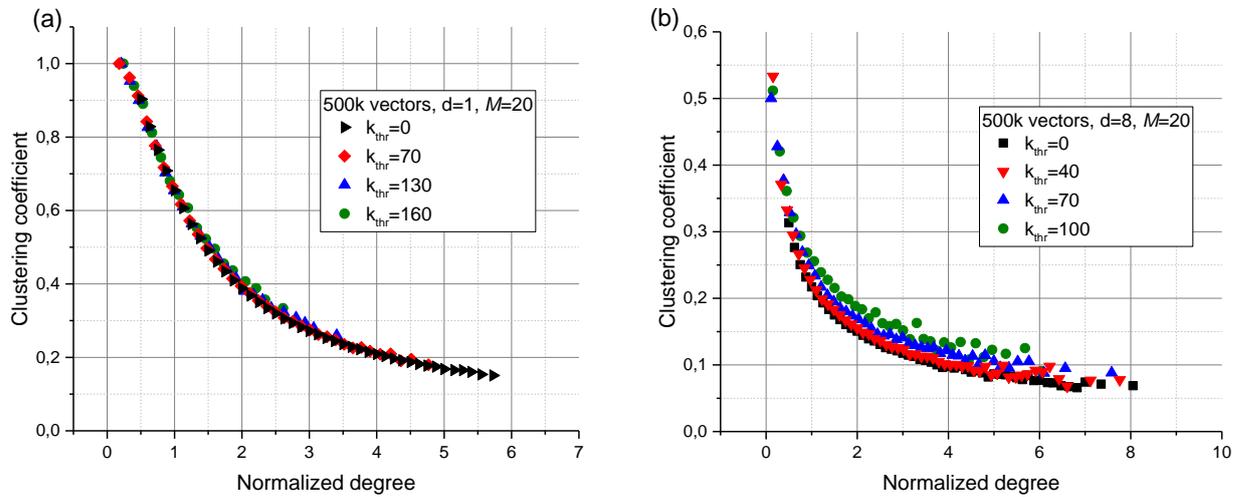

**Fig. S3.** Normalized average clustering coefficient distribution in the network for different values of $k_{thr}$ and dimensionality.

### 3. Average nearest neighbor degree

The plot shows average nearest neighbor degree distribution demonstrating that GH networks have assortative degree mixing.

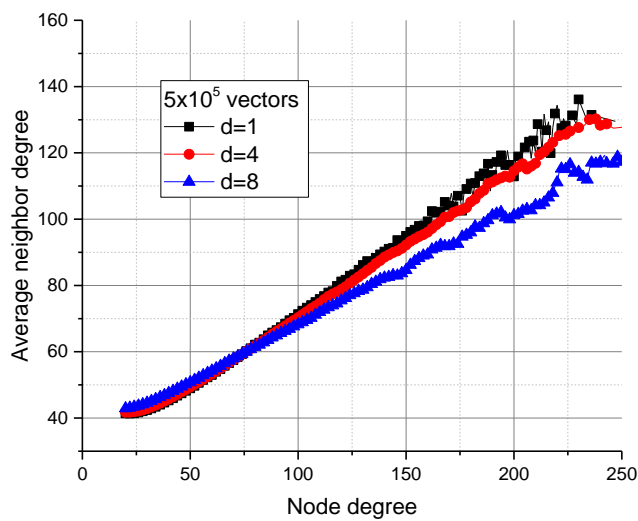

**Fig. S4.** Nearest neighbors degree distribution for vectors with different dimensionality.

### 4. Rich club coefficient

Figure S5 shows an exponential rise of the rich club coefficient $\phi(k) = 2E_{>k} / (N_{>k}(N_{>k} - 1))$ together with a rich club coefficient of a random network with the same degree distribution (for the case of d=8), which demonstrates that GH networks have rich clubs.

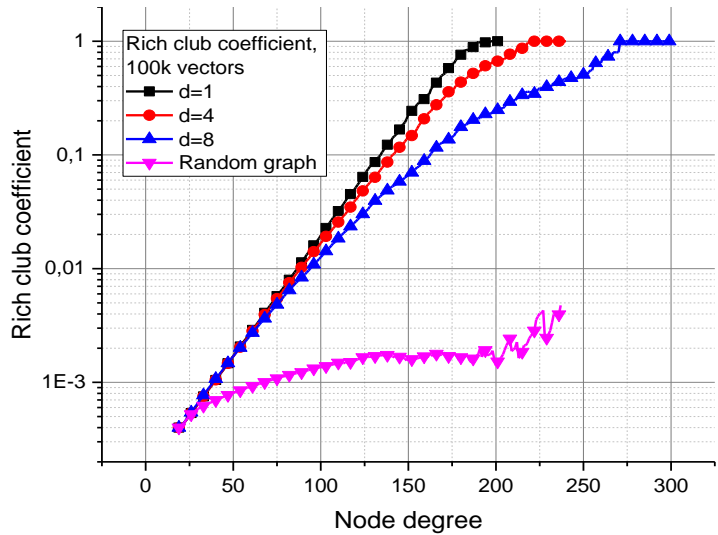

**Fig. S5.** Rich club coefficient of GH networks for vector data of different dimensionality

**5. Upper bounds of average greedy algorithm hops for the GH model**

Suppose we have a GH network which contains a perfect Delaunay graph at every step, rich club navigation feature and has an average greedy algorithm hop count *H*. We can show that by doubling the number of the elements the average greedy path increases no more than by adding a constant.

If the start and target nodes are from the rich club (first half of the network), the average hop count does not increase as it has been shown in the manuscript. If the greedy algorithm starts a search for a distant target from a newly added element it has at least 1/2 probability that the next selected element is from the rich club (since a half of the elements is from the rich club), thus reaching the rich club on average in two steps. Next, the greedy algorithm needs on average no more than *H* steps to get to a rich club element for which the Voronoi region has the query. For a case when the destination node is from the rich club the average number of hops is thus *H*+2. In the opposite case the search ends on average in two additional steps, because the probability that next node is from the rich club is at least 1/2 and since we have already visited the rich club node that is closest to the query, this cannot happen. Thus the average number of hops in this case is *H*+4. By concerning the last case (the target – newly added element and start element is from the rich club) we get an average *H*+2 hops. Thus the upper hop bound scales as $2\log_2(N)$ proving GH networks have a logarithmic scaling of the greedy search hops.

**References**

1. Boguna M, Krioukov D, Claffy KC (2009) Navigability of complex networks. Nature Physics 5: 74-80.
2. Krioukov D, Papadopoulos F, Kitsak M, Vahdat A, Boguná M (2010) Hyperbolic geometry of complex networks. Physical Review E 82: 036106.
3. Serrano MA, Krioukov D, Boguná M (2008) Self-similarity of complex networks and hidden metric spaces. Physical Review Letters 100: 078701.